# Observation of magnetic domains in uniaxial magnets via small-angle electron diffraction and Foucault imaging

Hiroshi Nakajima[1], Atsuhiro Kotani[1], Ken Harada[1,2], and Shigeo Mori[1*]

[1]*Department of Materials Science, Osaka Prefecture University, Sakai, Osaka 599-8531, Japan.*
[2]*Center for Emergent Matter Science, the Institute of Physical and Chemical Research (RIKEN), Hatoyama, Saitama 350-0395, Japan*

*E-mail: mori@mtr.osakafu-u.ac.jp

Observation of magnetic domains is important in understanding the magnetic properties of magnetic materials and devices. In this study, we report that the magnetic domains of *M*-type hexaferrites with uniaxial anisotropy can be visualized via small-angle electron diffraction and Foucault imaging. The position of the diffraction pattern spots has the same period as that of magnetic domains in a Sc-substituted hexaferrite ($BaFe_{12-x-\delta}Sc_xMg_\delta O_{19}$). Conversely, the spots were observed four times longer than the period of magnetic domains in hexaferrite without substitution ($BaFe_{12}O_{19}$), demonstrating the long-range order of the Bloch walls. When the specimen was tilted, the magnetic deflection effect, as well as the periodic spots of magnetic domains, occurred. Thus, we were able to visualize the magnetic domains with different magnetization directions and domain orientations by selecting deflection spots. The results indicate that the technique utilized in this study is useful in observing the magnetic materials with uniaxial anisotropy.

Keywords: Small-angle electron diffraction, Foucault imaging, magnetic domains, hexagonal ferrites, Bloch wall

# 1. Introduction

Ferromagnetic materials have been extensively used in industrial appliances, such as motors, inductances, and memory devices. Recently, hexagonal ferrites (hexaferrites), Sm–Co, and Nd–Fe–B magnets have been extensively investigated to improve the magnetic properties of electric appliances because these materials exhibit high magnetizations, uniaxial anisotropy, and transition temperatures above room temperature.[1,2] The magnetic microstructures should be significantly examined to evaluate the magnetic properties of these magnets.

To observe the magnetic domains in the magnets, several methods have been used, such as x-ray magnetic circular dichroism, magnetic force microscopy, spin-polarized scanning electron microscopy, and Lorentz remission electron microscopy (Lorentz microscopy hereinafter). Among them, the Lorentz microscopy has high spatial resolution and availability for controllable experimental conditions, such as temperature, external magnetic field, current induction, and stress.[3–6] It mainly involves five techniques, namely, Fresnel imaging, Foucault imaging, electron holography (EH), differential phase contrast imaging (DPC), and small-angle electron diffraction (SmAED).[7,8] The first two methods provide qualitative descriptions on the spatial variation of magnetization, whereas the rest supply quantitative information on magnetic domain structures. Fresnel imaging is a method used to depict domain walls by defocusing the imaging lens, whereas Foucault imaging is utilized to visualize magnetic domains by selecting magnetically deflected spots with an aperture. Unlike Fresnel imaging, Foucault imaging has been used less frequently because performing Fresnel imaging by defocusing the imaging lens is easy.

Recently, optical systems employed in performing Foucault imaging combined with SmAED were constructed using a conventional transmission electron microscope.[9–12] These optical systems allow the observation of magnetically deflected spots and the visualization of domains that cause the corresponding spots. Foucault imaging is advantageous for magnetic domain observations because magnetic domains with particular magnetization orientations can be visualized. The SmAED is useful in quantitatively analyzing magnetization and domain periods and investigating the symmetry of magnetic structures.[13–16] Unlike other quantitative methods (e.g., EH and DPC), it does not require special equipment, such as electron biprism and segmented detector. In addition, information obtained using the SmAED is collected from a low-angle range; thus, the measured values are not affected by diffraction effects, such as bend contours. However, the SmAED cannot provide real-space images of magnetic domains. Therefore, this electron diffraction is useful when a combination of Fresnel imaging, SmAED, and Foucault imaging is utilized in Lorentz microscopy observations.

In recent observations that employ SmAED, its applications were only restricted to 180° domains in ferromagnetic materials[9,17,18] and helical domains in chiral magnets[19,20]. As mentioned previously, magnetic materials with uniaxial anisotropy, such as hexaferrites and Nd–Fe–B magnets, are important for industrial appliances. However, observing magnetic domains with uniaxial anisotropy via Foucault imaging is difficult because the Lorenz force is not caused when the incident beam is parallel to the magnetization.

In this paper, we report the results of observing magnetic domains with uniaxial anisotropy in Sc-substituted $M$-type hexaferrites ($BaFe_{12-x-\delta}Sc_xMg_\delta O_{19}$)[21,22] via the SmAED and Foucault imaging. We obtained the SmAED patterns demonstrating the periodic spots in hexaferrites. The periodic spots in the patterns were generated from the Bloch wall when the incident beam was parallel to the easy axis of the magnetization. In $BaFe_{12}O_{19}$ without substitution, the width estimated from the periodic position was four times longer than that of magnetic domains because the Bloch walls had the ↑↑↓↓-type order. By tilting the specimen, the magnetic deflection from magnetic domains was observed apart from the periodic spots in SmAED patterns. Thus, we were able to visualize the magnetic domains with different magnetization directions and domain orientations using the Foucault images. This paper discloses the useful technique for observing magnetic domains magnetized parallel to the observation direction through the SmAED and Foucault imaging.

## 2. Experimental methods

Figure 1 shows the schematic of the optical system by (a) the SmAED and (b) Foucault imaging.[10,11] In the optical system, the objective lens was switched off to ensure that an external magnetic field was not applied to the specimen. External magnetic fields can be applied up to 200 mT by adjusting the currents of the objective lens and objective mini lens in this optical system. Condenser lenses were strongly excited also to ensure that the crossover size became small and that the divergence angle of the irradiation beam was small. Then, the objective mini lens and condenser lenses were adjusted to locate the crossover at a selected-area (SA) aperture plane. The SA aperture functions as an angle-limiting aperture. By switching from diffraction to Foucault modes, the intermediate lens I was less excited after selecting the diffraction spots. By selecting magnetically deflected spots, the magnetic domains that cause the corresponding spots can be visualized in bright contrast (Foucault imaging). Moreover, the objective aperture can be used to limit the specimen area for SmAED. The Fresnel imaging can also be performed by rendering the intermediate lens I out of focus. We utilized transmission electron microscopes (JEM-2010 and JEM-2100F, JEOL Co. Ltd.) to observe magnetic materials with uniaxial anisotropy. The accelerating voltage was 200 kV. The specimens were tilted by using a goniometer equipped with the microscopes. The optical system was constructed by using free lens control. The resolution of SmAED was better than 0.5 µrad in the optical system because the diffraction spots of 0.5 µrad were resolved.

The observation method was demonstrated in magnetic domains of Sc-substituted *M*-type barium hexaferrites $BaFe_{12-x-\delta}Sc_xMg_\delta O_{19}$ ($x = 1.6, \delta = 0.05$) (BFSMO hereinafter) and $BaFe_{12}O_{19}$ without substitution.[21,22] BFSMO and $BaFe_{12}O_{19}$ have the ferrimagnetic state whose net magnetic moments point parallel to the *c* axis at room temperature. The specimens for the observation were prepared as follows. The polycrystals were synthesized on the basis of the solid-state reaction from high-purity (4N up) powders of $BaCO_3$, $Fe_2O_3$, MgO, and $Sc_2O_3$. The powders were weighted on the basis of the required chemical compositions and were mixed in a mortar. Then, the mixed powders were calcined at 980 °C for 20 h in air. Thereafter, single crystals of the *M*-type barium hexaferrites were grown through a floating-zone method in the $O_2$ flow.[23] The single crystals were cleaved in the *c* plane. Thin specimens whose surface was perpendicular to the *c* axis were prepared via Ar ion milling technique for Lorentz microscopy observations. Moreover, the specimen thickness was less than 100 nm.

## 3. Results and discussion

Figure 2(a) shows the Fresnel image of BFSMO at the *c* plane without tilting the specimen. Owing to the Lorentz deflection, Bloch walls were visualized as pairs of bright and dark lines [Fig. 2(b)]. Figure 2(c) illustrates a SmAED pattern of BFSMO obtained with the incident beam parallel to the *c* axis. This pattern comprises spots and streaks. These spots are aligned in two directions that correspond to the V-shaped magnetic domains. The spots were due to the period of the magnetic domain walls because the distance of approximately $d = 180$ nm corresponds to the spot interval of $\theta = \lambda/d = 13.4$ µrad, where $\lambda = 2.51$ pm is the wavelength of the electron. The streaks were caused by the Bloch walls wherein magnetization continuously rotates.

Subsequently, we tilted the specimen by 30° to investigate the change in the diffraction pattern. Figure 3(a) shows the Fresnel image of BFSMO at the tilt angle of 30°. Unlike in Fig. 2(b), the domain wall in Fig. 3(b) is represented with a bright or dark line. As shown in the SmAED pattern of Fig. 3(c), the intensity of the direct beam spot weakens when the specimen was tilted. Before tilting the specimen, the magnetizations in magnetic domains were "up" (antiparallel) or "down" (parallel to the electron beam) and did not have the in-plane component, as illustrated in Fig. 2(d). Thus, the electron that passes these domains were not affected by the Lorentz force, resulting in causing the direct beam spot indicated by a white arrow in Fig. 2(c). However, when the specimen was tilted, these domains had in-plane components [see Fig. 3(d)]; consequently, the direct beam spot disappeared by deflecting the electrons, as observed in Fig. 3(c). As shown in Figs. 2 and 3, the magnetization direction parallel to the electron beam can be deduced from the SmAED patterns by tilting the specimen. The Lorentz deflection increased the intensity of the spots indicated by A, B, C, and D in Fig. 3(c); the periodic patterns were superimposed on the Lorentz deflection. The interval of the periodic spots at 6.8 μrad corresponds to twice longer than the Bloch wall period (180 nm). The reason for the double period is that a pair of up- and down-magnetized domains is the periodic structure for the incident electrons when the specimen was tilted [denoted by the blue marks in Figs. 2(d) and 3(d)]. A similar phenomenon was also observed in another ferromagnet.[24]

We observed $BaFe_{12}O_{19}$ to compare the results in BFSMO. Figure 4(a) shows the Fresnel image of $BaFe_{12}O_{19}$. The domain width was approximately 250 nm. In the SmAED pattern shown in Fig. 4(b), the periodic spots were observed at 2.6 μrad, which corresponds to $(980\ nm)^{-1}$. Accordingly, the period evaluated from the spots is four times longer than that of the domain width. When observing the contrast at the Bloch walls in the Fresnel image, we identified that the magnetizations with right–right–left–left (↑↑↓↓) directions were repeated at the Bloch walls (denoted by the arrows). These results suggest that the periodic spots were caused by the Bloch walls. This discussion is also applicable to the periodic spots in BFSMO because the period of Bloch walls is the same as the domain width. When the $BaFe_{12}O_{19}$ specimen was tilted, the Fresnel image [Fig. 4(c)] had bright or dark contrast at the Bloch walls. This observation is similar to that of BFSMO. The spots in the SmAED of Fig. 4(d) are 5.3 μrad, which corresponds to approximately twice of the periodicity (250 nm). Therefore, the periodic spots were due to the magnetic domains, similar to the case of BFSMO, when the specimen was tilted. In contrast, the periodic spots were generated by the Bloch walls before tilting the specimens.

The long-range (↑↑↓↓) order of the Bloch walls in $BaFe_{12}O_{19}$ is difficult to be noticed from the Fresnel image [Fig. 4(a)]. As demonstrated in Fig. 4, SmAED has the advantage to display unnoticeable periodic magnetic structures and to determine the magnetization directions. The one possible origin of the order of Bloch walls is assumed to be the balance between the dipole–dipole interaction and stray field energy.[25,26] The dipole–dipole interaction is long range and expressed by the following equation:

$$E_{\text{dipole}} = \frac{\mu_0 M_s^2}{4\pi} \int d\bm{r} \int d\bm{r}' \left\{ \frac{\bm{m}(\bm{r}) \cdot \bm{m}(\bm{r}')}{|\bm{r}-\bm{r}'|^3} + \frac{\bm{m}(\bm{r}) \cdot (\bm{r}-\bm{r}')\, \bm{m}(\bm{r}') \cdot (\bm{r}-\bm{r}')}{|\bm{r}-\bm{r}'|^5} \right\}. \quad (1)$$

Here, $M_s$ is saturation magnetization, $\mu_0$ is vacuum permeability, and $\bm{m}$ is magnetization vector. The stray field energy represents the magnetostatic energy generated from the specimen to vacuum:

$$E_s = \frac{1}{2}\mu_0 \int H_d^2\, d\bm{r}, \quad (2)$$

where $H_d$ is stray field. The dipole–dipole interaction favors the parallel magnetization. In contrast, the stray field energy forces the magnetic domains to form flux-closed domain structures (180° or 90° domains) to avoid the increase in the stray field. When only the dipole–dipole interaction is considered, the magnetizations at the adjacent Bloch walls point the same direction. However, the parallel magnetizations increase the stray field energy owing to the magnetic field in vacuum. Thus, the antiparallel magnetizations were also formed to reduce the stray field energy, resulting in the formation of the long-range (↑↑↓↓) order of Bloch walls. For BFSMO that showed the parallel Bloch walls, possible origins of the Bloch wall structures are the in-plane demagnetization field due to the wedge-shaped specimen and residual magnetic field from the electron microscope. However, these effects can be ignored when the magnetocrystalline anisotropy in $BaFe_{12}O_{19}$ is large. It is revealed that the magnetocrystalline anisotropy can be reduced by the Sc substitution and the helicity (clockwise or counterclockwise) of the Bloch walls was randomly distributed in BFSMO[22], supporting the discussions that in-plane magnetic fields aligned with the Bloch walls.

Furthermore, we performed the Foucault imaging in BFSMO to demonstrate that the present optical system is useful for the magnetic domain observation. By selecting the deflected spots (A and B), we were able to obtain a Foucault image that visualizes the up-magnetized domains as bright contrast in a large area (Fig. 5). Although the Fresnel image [Fig. 2(a)] illustrates the Bloch wall positions, the Foucault image shows the magnetic domains with the same magnetization directions.

In addition, we selectively visualized the magnetic domains with various magnetization directions and domain orientations. Figures 6(a)–6(f) show the contrast change in the Foucault imaging using a single or a pair of deflected spots. As presented in Fig. 6(a), where spots A and B were used for visualization, bright and dark lines were alternately depicted, corresponding to up- and down-magnetized domains, respectively. Unlike in Fig. 6(a), the contrast of the domains in Fig. 6(b) is interchanged, and the down-magnetized domains have bright contrast if spots C and D were selected. As presented in Fig. 6(c), we selected spots A and C from the up-magnetized domains in the $x$ directions and the down-magnetized domains in the $y$ directions. Consequently, the bright and dark contrast is interchanged where the domains change the directions. Meanwhile, their contrast was reversed when spots B and D were selected, as demonstrated in Fig. 6(d). The only up-magnetized domains in the $x$ directions were bright, and the other regions were dark when spot A was used in Fig. 6(e). Similarly, the up-magnetized domains in the $y$ directions were bright, as shown in Fig. 6(f) in which spot B was selected. As shown in Fig. 6, the Foucault images can depict the domains with the selected magnetization direction and domain orientations using the present method.

## 4. Conclusions

We demonstrated that the magnetic domains magnetized parallel to the incident beam direction can be visualized via Foucault imaging and SmAED. SmAED patterns showed periodic spots that correspond to the Bloch wall periods. The ↑↑↓↓-type Bloch walls were revealed on the basis of the SmAED pattern. By tilting the specimen, the magnetization direction was deduced from the SmAED patterns. Furthermore, the magnetic deflection from magnetic domains appeared, as well as the periodic spots after tilting the specimens. The results showed that the magnetic domains with various magnetization directions and domain orientations can be visualized by selecting the corresponding spots as Foucault images. These results will help in observing and analyzing the magnetic domains via Foucault imaging and SmAED.

## Acknowledgments

This work was partially supported by JSPS KAKENHI (Nos. 16H03833 and 15K13306) and a grant from the Murata Foundation.

## Figure Captions

**FIG. 1.** Schematics of the optical system for (a) small-angle electron diffraction (SmAED) and (b) Foucault imaging. The intermediate lens I is adjusted when the SmAED mode is switched to the Foucault mode. In the Foucault mode, the magnetically deflected spots are selected by using a selected-area aperture to visualize the magnetic domains.

**FIG. 2.** (a) Fresnel image of $BaFe_{12-x-\delta}Sc_xMg_\delta O_{19}$ ($x = 1.6, \delta = 0.05$) (BFSMO) at the $c$ plane. The defocus is $\Delta f = -1.5$ μm (underfocus). (b) Intensity profile along X1–Y1 in (a). Bloch walls are visualized as pairs of bright and dark lines. (c) SmAED pattern performed at a camera length of 380 m. The arrows show the positions of the direct beam. (d) Schematic of the magnetic domains in (a). The red arrows show the magnetization, whereas the blue marks represent the periods of magnetic domains for the electron.

**FIG. 3.** (a) Fresnel image of BFSMO when the specimen was tilted to 30°. (b) Intensity profile along X2–Y2 in (a). (c) SmAED pattern at the tilt angle of 30°. The letters (A–D) denote the magnetic deflection spots used in Foucault imaging in Fig. 6. A Bloch wall is depicted as bright or dark line. (d) Schematic of the magnetic domains in (a).

**FIG. 4.** (a) Fresnel image of $BaFe_{12}O_{19}$ without tilting the specimen. The arrows show the directions of the magnetization at the Bloch walls. (b) SmAED pattern without the specimen tilt. (c) Fresnel image at the tilt angle of 30°. (d) SmAED pattern at the tilt angle of 30°. In the Fresnel images, the defocus was $\Delta f = -2.0$ μm (underfocus). The arrows represent the magnetization directions at (a) domain walls and (b) magnetic domains. The SmAED patterns were obtained at a camera length of 300 m.

**FIG. 5.** Foucault image of BFSMO. The image was obtained by selecting the A and B spots of Fig. 3(c).

**FIG. 6.** Foucault images when selecting the magnetic deflection spots A–D in Fig. 3(c). The used spots are A and B in (a), C and D in (b), A and C in (c), B and D in (d), A in (e), and B in (f). In (a), the up-magnetized domains are bright, whereas the down-magnetized ones are dark. In (b), the down-magnetized domains have bright contrast and its contrast is reversed compared with (a). (c) The up-magnetized domains along the *y* direction and down-magnetized domains along the *x* direction are bright. Therefore, the bright and dark contrast is reversed where the domains bend. Meanwhile, the down-magnetized domains along the *y* direction and up-magnetized domains along the *x* direction in (d) are bright. In (e), only the up-magnetized domains along the *x* direction are bright. In (f), only the up-magnetized domains along the *y* direction are bright.

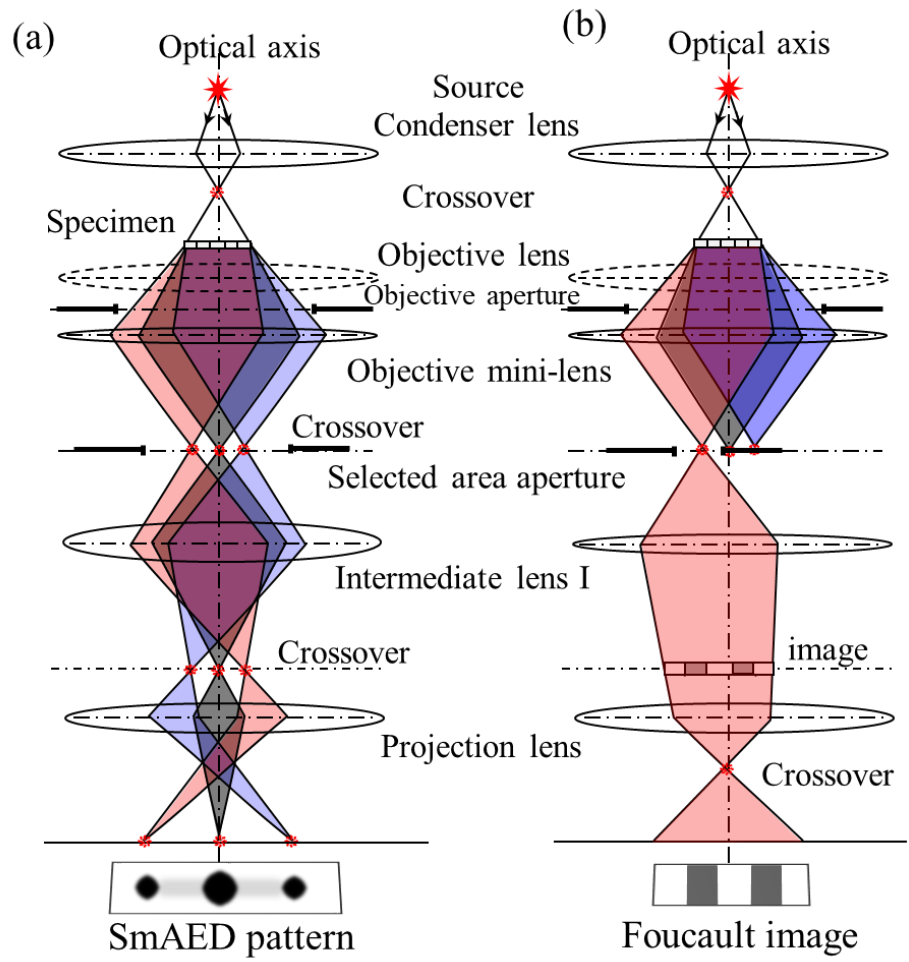

Fig. 1.

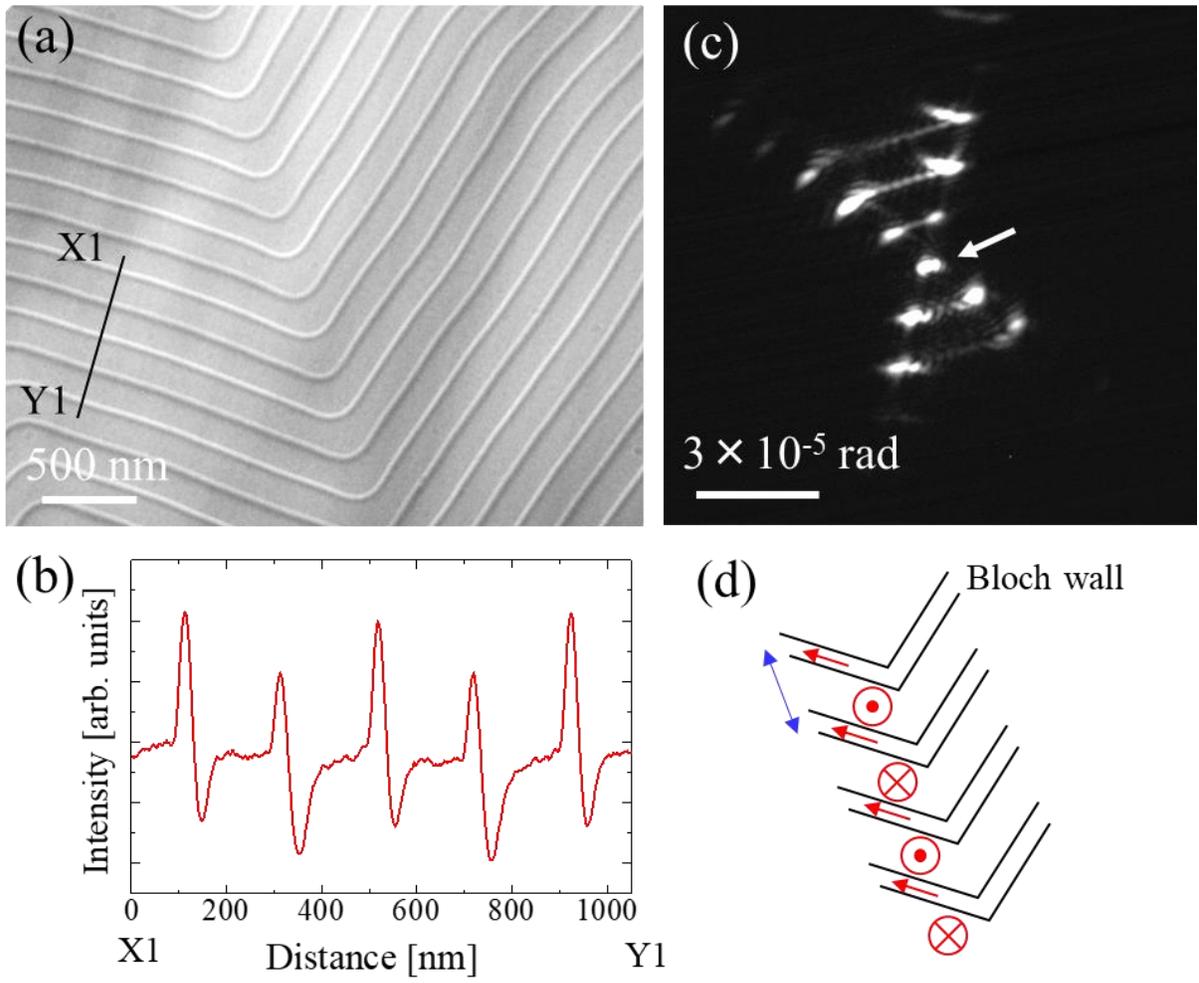

FIG. 2.

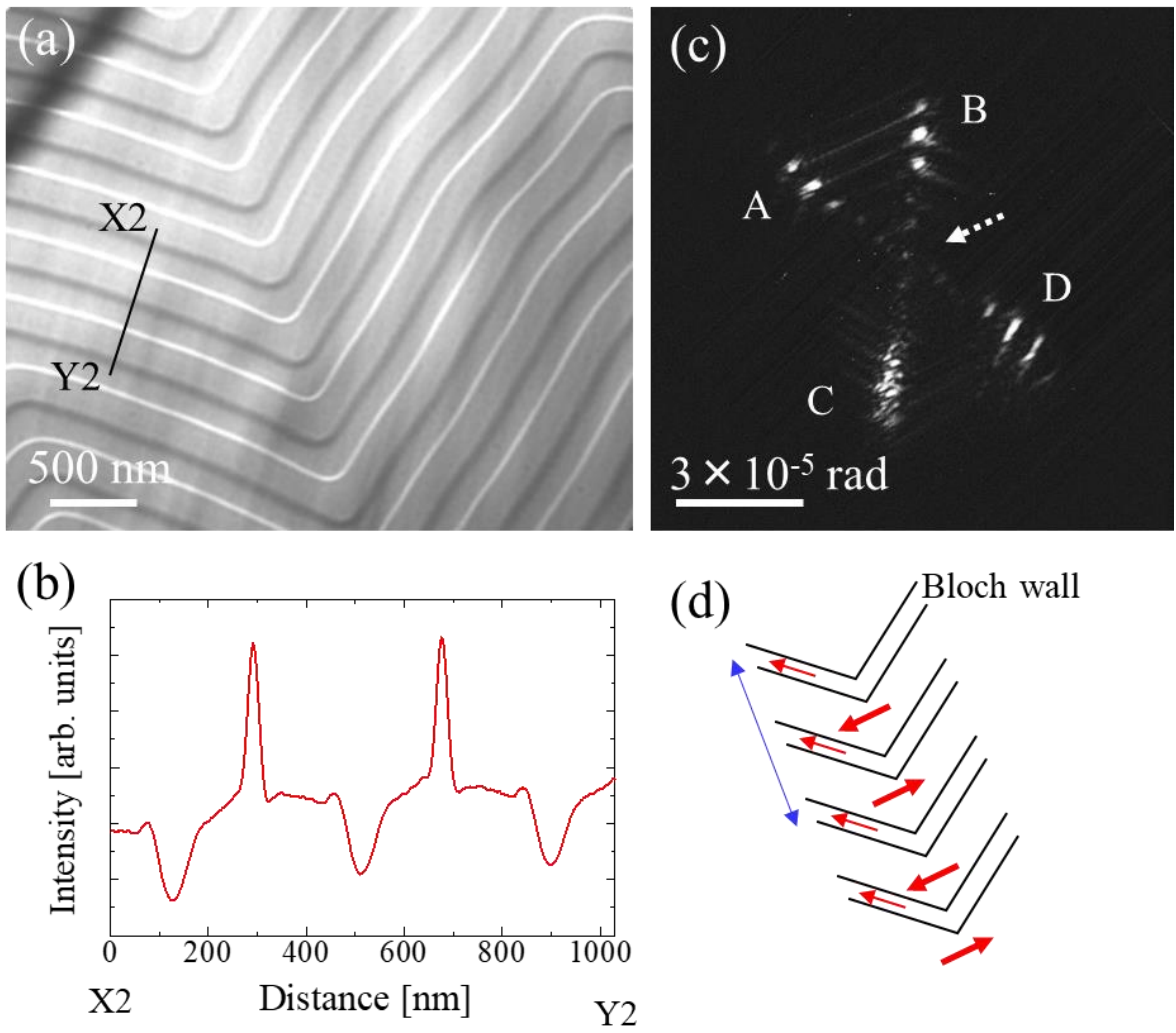

FIG. 3.

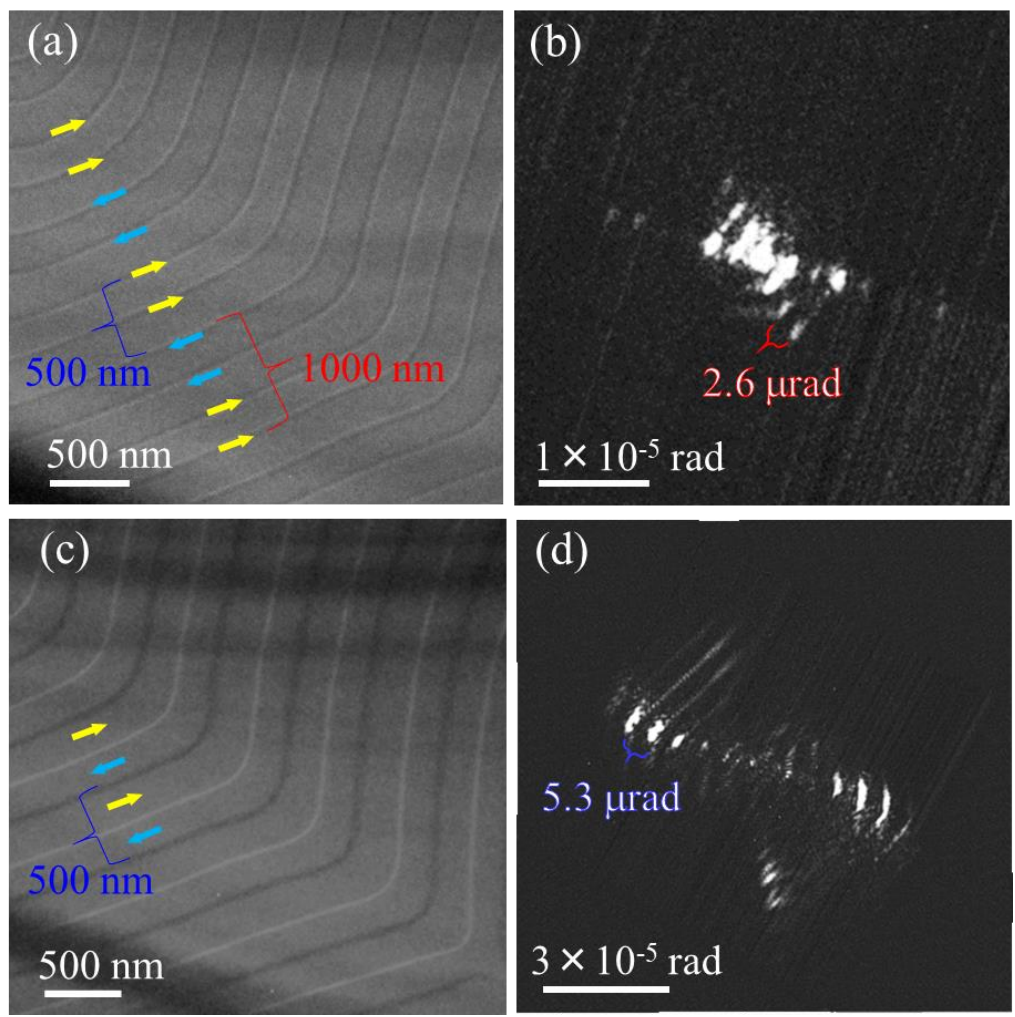

FIG. 4.

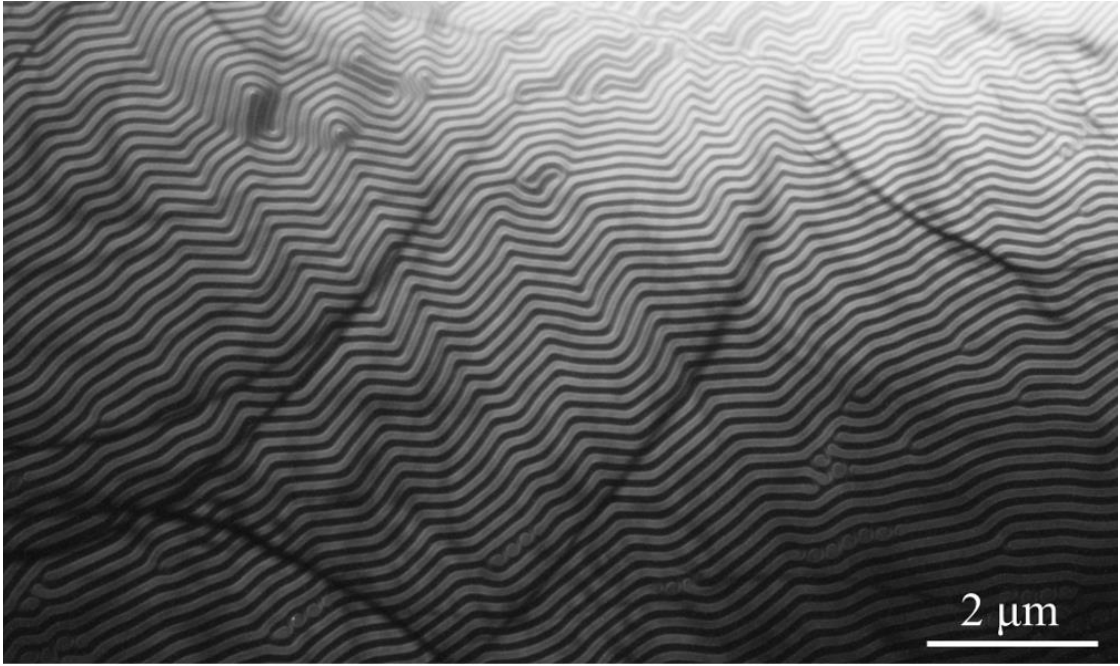

FIG. 5.

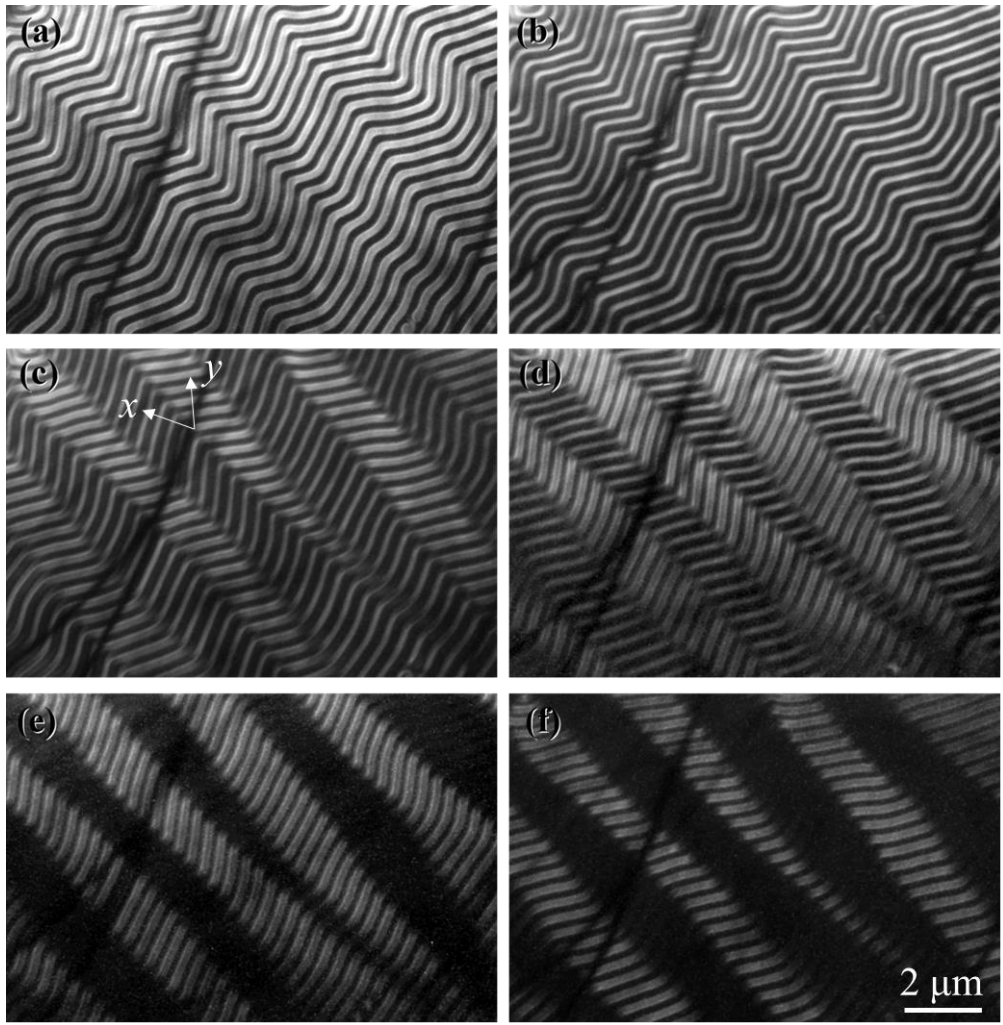

FIG. 6.